\documentclass[12pt]{iopart}

\usepackage{psfig}
\usepackage{bm}

\begin{document}

\newcommand{\suin}{\sum \hspace*{-5mm} \int\limits}

\title[Influence of electron-ion collisions on Coulomb
crystallization...]{Influence of electron-ion collisions on Coulomb
crystallization of ultracold neutral plasmas}
\author{T Pohl, T Pattard and JM Rost}
\address{Max Planck Institute for the Physics of Complex Systems,
N\"othnitzer Stra{\ss}e 38, D-01187 Dresden, Germany}
\ead{tpohl@mpipks-dresden.mpg.de}

\begin{abstract}
While ion heating by elastic electron-ion collisions may be neglected for a
description of the evolution of freely expanding ultracold neutral plasmas, the
situation is different in scenarios where the ions are laser-cooled during the
system evolution. We show that electron-ion collisions in laser-cooled
plasmas influence the ionic temperature, decreasing the degree of correlation
obtainable in such systems. However, taking into account the collisions
increases the ion temperature much less than what would be estimated based
on static plasma clouds neglecting the plasma expansion. The latter leads to
both adiabatic cooling of the ions as well as, more importantly, a rapid
decrease of the collisional heating rate.
\end{abstract}

\pacs{32.80.Pj,52.27.Gr}
\submitto{\jpb}

\maketitle

Recently, the field of ultracold ($T \ll 1$K) neutral plasmas and Rydberg gases
has attracted attention both experimentally \cite{Kil99,Rob00,Rai01} and
theoretically \cite{Mur01,Rob02,Kuz02,Maz02,Tka01,PPR03}. One of the
motivations of
the experiments \cite{Kil99,Kil04} is the creation of a so-called strongly
coupled plasma, where the {\sl Coulomb coupling parameter} $\Gamma_{\rm i} =
e^2/(a k_{\rm B} T_{\rm i})$ is much larger than unity (where $a=(4 \pi
\rho_{\rm i} /3)^{-1/3}$ is the Wigner-Seitz radius and $T_{\rm i}$ is the
temperature of the ions). In such a case, interesting ordering effects such as
Coulomb crystallization into cubic or shell-structure lattices can be observed.
In the experiments \cite{Kil99} strongly coupled ions can not be observed
since the temperature of the initially cold ions rises quickly on the
timescale of the inverse ionic plasma frequency
$\omega_{\rm{p,i}}^{-1}=\sqrt{m_{\rm{i}}/\left(4\pi e^2\rho_{\rm{i}}\right)}$ due to
disorder-induced heating \cite{Mur01,Kuz02,Maz02}. Different ways to overcome
this heating effect have been proposed
\cite{Mur01,Ger03,Kuz02b,Kil03,PPR04a,PPR04b}. In \cite{Kuz02b,Kil03}, e.g.,
it has been suggested that
continuous laser cooling during the plasma expansion may considerably increase
the achievable Coulomb coupling parameter of the ions.

Using a hybrid molecular dynamics approach \cite{PPR04c},
we have predicted that Coulomb crystallization in an unconfined ultracold
neutral plasma can indeed be obtained if the plasma ions are laser-cooled
during the evolution of the system \cite{PPR04a}. In this theoretical
approach, the electronic component of the plasma is treated as a fluid
while ions and recombined atoms are described on a full molecular dynamics
level. More precisely, an adiabatic approximation is made for the electrons,
which are assumed to be distributed according to the mean-field potential
generated by ions and electrons, calculated in a self-consistent way. Their
velocity distribution is assumed to be of Michie-King-type \cite{Kin66} which
takes into account deviations of the quasi-equilibrium state from a
Maxwell-Boltzmann distribution due to the finite depth of the potential well
generated by the ions. The latter leads to an evaporation of a fraction
of the electrons in the initial stage immediately after the plasma formation,
which is taken into account using the results of \cite{Kil99}.
The ions, on the other hand, are propagated under the influence of all other
ions and the electronic mean-field. Finally, inelastic collisions, i.e.\
three-body
recombination, electron-impact ionization and electron-impact (de)excitation,
are included on the basis of a Monte Carlo treatment. {\em Elastic}
electron-ion collisions, however, are
neglected in this approach, as in other approaches for the description of
ultracold plasmas \cite{Rob03}.

This neglect of elastic electron-ion collisions in the dynamics is well
justified for the description of freely expanding ultracold plasmas as created
in \cite{Kil99} because of a clear separation of timescales. Due to the
large mass ratio between ions and electrons, the time necessary for
equilibration of the electron and ion temperature is typically of the order of
a few milliseconds, while the plasma expansion takes place on a microsecond
timescale. Hence, elastic electron-ion collisions only increase the ion
temperature by some milli-Kelvin during the experimental observation time
\cite{Kil99}. This
amount of heating is negligible compared to the initial temperature increase
due to disorder-induced heating, which rises the ion temperature to
about one Kelvin. Hence, the time evolution of the ion temperature which enters
$\Gamma_{\rm i}$ and determines the degree of
coupling of the ions is mainly determined by this disorder-induced heating.

However, this situation changes in the scenario of \cite{PPR04a}, where
additional laser cooling of the ions compensates the disorder-induced heating
and keeps the ionic temperature on a milli-Kelvin level. In this case, the ion
temperature is not only increased by electron-ion collisions, but is also
driven to the Doppler temperature, i.e.\ the limiting temperature $T_{\rm c}$
for laser cooling, due to the
coupling to the
radiation field. The final ion temperature is now determined by the
balance between the collisional heating rate and the laser-cooling rate
\cite{Kuz02b,Kil03}. Since $T_{\rm c}$ is typically of the order of one
milli-Kelvin, i.e.\ of the same order of magnitude as the amount of collisional
heating, the ion temperature may considerably increase through electron-ion
collisions. Thus,
electron-ion collisions could significantly decrease the Coulomb coupling
parameter achievable by laser-cooling of expanding ultracold neutral
plasmas. In the following, the influence of such collisions on the onset
of Coulomb crystallization during the plasma expansion is critically reassessed.

The system under study is the same as that of \cite{PPR04a}, namely an
unconfined, ultracold neutral plasma under the additional influence of a
cooling laser. The plasma is treated on the basis of
the hybrid-MD method outlined above and described in detail in \cite{PPR04c}.
Laser cooling is modelled by adding a Langevin force, ${\bf{F}}_{\rm cool}
=-m_{\rm{i}}\beta{\bf{v}}_{\rm{i}}+\sqrt{2\beta k_{\rm{B}}T_{\rm{c}}m_{\rm{i}}}{\bm{\xi}}$, to the
ion equation of motion, where ${\bf{v}}_{\rm{i}}$ is the ion velocity,
${\bm{\xi}}$ is a stochastic variable with $\left<{\bm{\xi}}\right>={\bf{0}}$, $\left<{\bm{\xi}}(t){\bm{\xi}}(t+\tau)\right>=3\delta(\tau)$ and the cooling rate $\beta$ and the
corresponding Doppler temperature $T_{\rm{c}}$ are determined by the
properties of the cooling laser \cite{Met99}. The weak electron
coupling, which justifies our hybrid treatment of the plasma dynamics,
also allows us to describe electron-ion collisions with a Boltzmann-type
collision operator, as we shall discuss below.
The implementation of electron-ion collisions is very similiar to the
treatment of electron-electron collisions used in \cite{Rob03} and has been
introduced before in \cite{Nan80}. For an ion at position $\mathbf{r}_{\rm i}$ with velocity $\mathbf{v}_{\rm
i}$, the rate of electron-ion collisions is
\begin{equation}
\label{rate}
K_{\rm eic} = \int d{\bf{v}}_{\rm{e}}d\Theta\:\sin\Theta\left( \frac{d
\sigma}{d\Theta} \right)_{\rm eff} \left|{\bf{v}}_{\rm{i}}-{\bf{v}}_{\rm{e}}
\right| f_{\rm{e}}({\bf{r}}_{\rm{i}},{\bf{p}}_{\rm{e}}) \; .
\end{equation}
Hence, $P_{\rm eic} = \Delta
t \; K_{\rm eic}$ is the probability that a collision will occur during a
timestep $\Delta t$. Numerically, the integration over the electronic
velocity distribution in equation (\ref{rate}) is done via a Monte Carlo
procedure. At each timestep, an electronic velocity $\mathbf{v}_{\rm e}$ is
chosen randomly according to a Maxwell-Boltzmann distribution with
temperature $T_{\rm e}$. The probability for an electron-ion collision is
then given by
\begin{equation}
\label{coprob}
P_{\rm eic} (\mathbf{v}_{\rm e}) = \Delta t\; \rho_{\rm{e}}({\bf{r}}_{\rm{i}})
v_{\rm{e}}\int_{-1}^{1}\left( \frac{d\sigma}{d\Theta} \right)_{\rm eff}
d(\cos\Theta)
=\Delta t\; \rho_{\rm{e}}({\bf{r}}_{\rm{i}})v_{\rm{e}}\frac{\pi
e^4}{m_{\rm{e}}^2v_{\rm{e}}^4}\left(\Lambda-1\right)\;,
\end{equation}
with $\ln \Lambda = \ln \left(\sqrt{3}/\Gamma_{\rm e}^{3/2} \right)$
the so-called Coulomb logarithm and
\begin{equation} \label{cs}
\left(\frac{d \sigma}{d \Theta} \right)_{\rm eff} = \left\{
\begin{array}{cl}
\left(\frac{d \sigma}{d \Theta} \right)_{\rm Coul} & \Theta \ge
\Theta_{\rm{min}}\\
0 & \Theta <\Theta_{\rm{min}}
\end{array}
\right.\;,
\end{equation}
where $\left(\frac{d \sigma}{d \Theta} \right)_{\rm Coul}$ is the Rutherford
cross section for Coulomb scattering \cite{Fri98} and $\Theta_{\rm{min}} = 2
\arcsin \Lambda^{-1/2}$ \cite{Li01}.
In the derivation of equation (\ref{coprob}), the approximation $m_{\rm e} /
m_{\rm i} \to 0, \mathbf{v}_{\rm i} \to 0$ has been made (see below).

In the Monte Carlo treatment outlined above, an electron-ion collision takes
place with a probability $P_{\rm eic} (\mathbf{v}_{\rm e})$. In this case,
a scattering angle $\Theta$ is determined according to the probability
measure $P(\Theta) \propto \left(\frac{d \sigma}{d \Theta} \right)_{\rm eff}$,
leading to the prescription
\begin{equation}
\cos \Theta = \frac{\zeta(\Lambda-1)-1}{\zeta(\Lambda-1)+1}\;,
\end{equation}
where $\zeta$ is a random number distributed uniformly in $[0,1]$. With this,
the change of ionic momentum is given by $\Delta {\bf{p}}_{\rm i} = m_{\rm{e}}v_{\rm{e}}\left(1-\cos\Theta\right){\bf{j}}$, where
$\mathbf{j}$ is a vector denoting the random direction of the momentum
transfer chosen uniformly on the unit sphere. Finally, energy conservation
is restored by adjusting the electronic temperature $T_{\rm e}$.

As mentioned above, the derivation of equation (\ref{coprob}) implies the limit
$m_{\rm e} / m_{\rm i} \to 0$. In this case, individual collision events
satisfy momentum conservation, but energy conservation is violated and has to
be corrected manually by adjusting the electron temperature. The error
introduced by this approximation can be shown to be of the order of the
temperature ratio $T_{\rm i} / T_{\rm e}$ \cite{unpub}. Hence, as long as the
ionic temperature is small compared to the electronic temperature, which is
the case over the whole time of the experiments under consideration, the
corresponding corrections can be neglected. The method described above thus
allows for an efficient simulation of electron-ion collisions without the need
for time-consuming transformations between laboratory and center-of-mass
coordinate systems.

\begin{figure}[tb]
\centerline{\psfig{figure=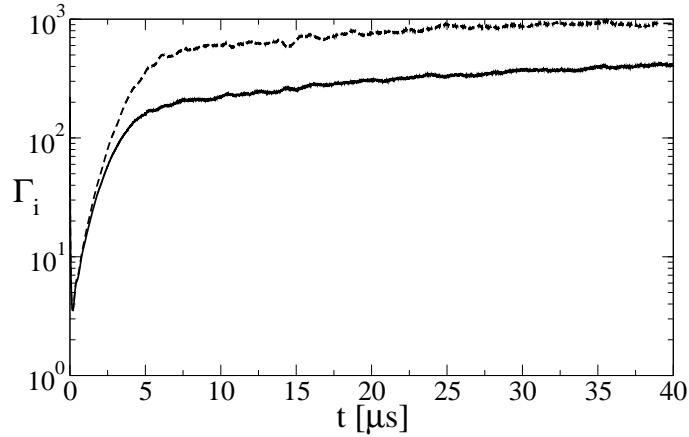,width=9cm}}
\caption{\label{fig1}
Time evolution of the ionic Coulomb coupling parameter with (solid) and without
(dashed) the inclusion of ion heating by electron-ion collisions for a plasma
of $20000$ Be ions with $\rho_{\rm i0} = 2.3 \times 10^8$
cm$^{-3}$, $\Gamma_{\rm{e0}}=0.05$, $\beta=0.10\:\omega_{\rm{p,i}}(t=0)$ and
$T_{\rm{c}}=1$mK.}
\end{figure}

In order to check the influence of elastic electron-ion collisions on the
plasma dynamics, we have simulated the expansion of a plasma with an initial
electronic Coulomb coupling parameter of $\Gamma_{\rm{e}}(t=0)=\Gamma_{\rm{e0}}=0.05$,
consisting of $20000$ Be ions at a density of $\rho_{\rm i}(t=0)=\rho_{\rm i0} = 2.3 \times 10^8$
cm$^{-3}$,
cooled with a damping rate of $\beta=0.10\:\omega_{\rm{p,i}}(t=0)$ and
a Doppler temperature of $1\:$mK. As can be seen in figure \ref{fig1},
\begin{figure}[bt]
\centerline{\psfig{figure=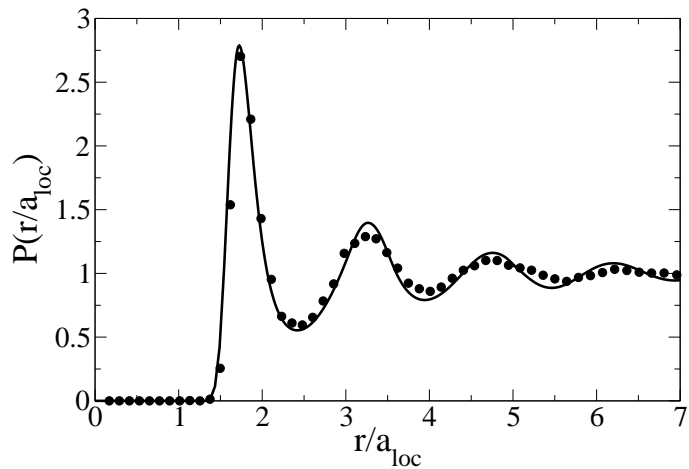,width=9cm}}
\caption{\label{fig2}
Distribution of scaled interionic distances after a time of $t = 40\:\mu$s,
compared to the calculated pair-correlation function of an OCP at $\Gamma_{\rm
i} = 400$. The initial-state parameters are the same as in figure \ref{fig1}.}
\end{figure}
the inclusion of elastic electron-ion collisions significantly reduces the
ionic coupling parameter which can be achieved in this case.
However, the value $\Gamma_{\rm{i}}$ obtained after several $\mu$s is still
considerably larger than the critical value $\Gamma_{\rm{i,c}}\approx 174$
\cite{Dub99} for crystallization. The good agreement found between the
distribution of interionic
distances, calculated as described in \cite{PPR04a}, and the pair-correlation
function of a one-component plasma at $\Gamma_{\rm{i}}=400$ \cite{Ng74} also
confirms that the system evolves well into the strongly coupled regime (figure
\ref{fig2}). We note here that Debye shielding of the ion-ion interaction,
which is not included in the present considerations, 
tends to increase the crystallization limit. However, as shown in \cite{Ham96},
even for a comparably large electron coupling of $\Gamma_{\rm{e}}=0.2$ the
crystallization limit is increased to $\Gamma_{\rm{i,c}}\approx198$
only, hence the inclusion of Debye screening does not significantly alter the
results obtained with the present model.

In fact, the simulation results show that the final $\Gamma_{\rm{i}}$ is
much larger than what one would expect from a static estimate by equating the
heating $\gamma_{\rm eic} T_{\rm e}$ resulting from electron-ion collisions to
the cooling $2 \beta (T_{\rm i} - T_{\rm c})$ of the laser cooling as done in
\cite{Kuz02b,Kil03}. From the procedure described above, $\gamma_{\rm{eic}}$
is obtained
as\footnote{Again, this implies the limit $T_{\rm i}/T_{\rm e} \to 0$ which
is well fulfilled over the whole timescale of the experiments as described
above.}
\begin{equation}
\gamma_{\rm eic} T_{\rm e} = \frac{2}{3k_{\rm{B}}}\int d{\bf{v}}_{\rm{e}}d\Theta\:\sin\Theta
\frac{m_{\rm e}v_{\rm e}^2}{2} \left(1 - \cos \Theta \right)^2 \left( \frac{d
\sigma}{d\Theta} \right)_{\rm eff} v_{\rm e} f_{\rm{e}}({\bf{r}}_{\rm{i}},{\bf{p}}_{\rm{e}}) \; ,
\end{equation}
leading to the Landau-Spitzer expression for the average heating rate
\cite{Ger02}
\begin{equation} \label{geic}
\gamma_{\rm eic}=\sqrt{\frac{2}{3\pi}}\frac{m_{\rm{e}}}{m_{\rm{i}}}
\Gamma_{\rm{e}}^{3/2}\omega_{\rm{p,e}}\ln\Lambda \; ,
\end{equation}
where $\omega_{\rm{p,e}}$ is the electronic plasma frequency. The validity and
extensions of the Landau-Spitzer formula, which was originally derived for
weakly coupled plasmas, have been discussed in several publications
\cite{Ger02,Lee84,Dha98,Haz01}. In \cite{Ger02}, ion heating by binary
collisions has been studied without
employing the cutoff-procedure equation (\ref{cs}). For  $\Gamma_{\rm{e}}<0.25$,
their numerical results are reproduced by equation (\ref{geic}) to within
$10\%$. Moreover, collective effects, which largely decrease the relaxation
rate at strong coupling \cite{Dha98}, were found to be negligible even
for arbitrarily strong ion-ion coupling, as long as the
electron-ion coupling is weak and the resonance of the ion excitation
spectrum lies far below that of the electrons \cite{Haz01}. Therefore, under
the present conditions, which correspond to the parameter regime studied in
\cite{Haz01}, the Monte Carlo treatment described above yields an adequate
description of the electron-ion temperature relaxation process.

Balancing collisional heating with the laser cooling one
might expect a quasi-equilibrium state with a final ion temperature of
\begin{equation}
\label{statictemp}
T_{\rm i} = \frac{\gamma_{\rm eic}}{2 \beta} T_{\rm e} + T_{\rm c} \; ,
\end{equation}
leading to a Coulomb coupling parameter
\begin{equation} \label{static}
\Gamma_{\rm{i}}=\sqrt{\frac{6\pi}{\Gamma_{\rm{e}}}}\frac{m_{\rm{i}}}{m_{\rm{e
}}} (\ln\Lambda)^{-1}\frac{\beta}{\omega_{\rm{p,e}}}
\end{equation} 
in the optimal limit that $T_{\rm c} \to 0$.
For the parameters used in figure \ref{fig1}, equation (\ref{static}) predicts
a final $\Gamma_{\rm{i}} \approx 50$, which is much smaller than
what is observed in the simulation. 

In order to trace the origin of this discrepancy, we have performed a second
set of simulations.
\begin{figure}[tb]
\centerline{\psfig{figure=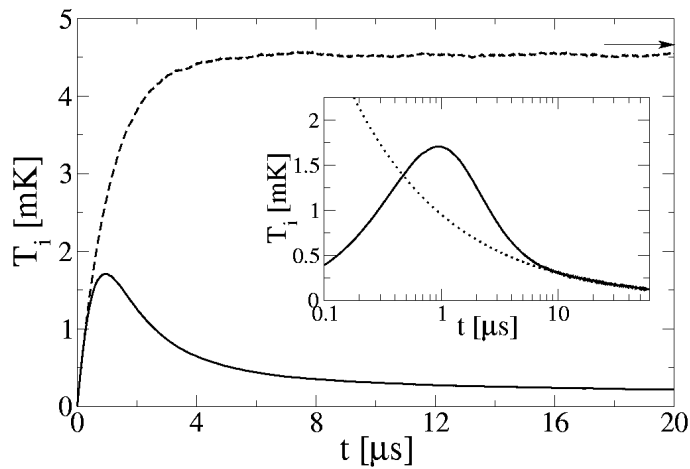,width=9cm}}
\caption{\label{fig3}
Time evolution of the ionic temperature for a plasma of 100000 ions with
$\rho_{\rm i0}=2 \times 10^7$ cm$^{-3}$, $\Gamma_{\rm{e0}}=0.08$ and 
$\beta=0.15\:\omega_{\rm{p,i}}(t=0)$, using different levels of approximation (see text).
The arrow shows the static estimate equation (\ref{statictemp}) for the
final ion temperature, the dotted line in the inset equation
(\ref{selfs}).}
\end{figure}
In one simulation,
we have set the Doppler temperature $T_{\rm c}$ equal to zero and have turned
off the initial correlation-induced heating of the ions by propagating them in
the framework of a particle-in-cell method \cite{Rob03}, i.e.\ on a mean-field level,
rather than by a full MD simulation. By doing this, only the competition
between the laser cooling, the heating by electron-ion collisions, and the
adiabatic expansion determines the evolution of the ion temperature. In the
second type of simulation, $T_{\rm c}$ is also set to zero and all particle
interactions except the
binary electron-ion collisions causing the ion heating are neglected, which
results in a basically static ionic density\footnote{The slow diffusion of
the ion distribution due to the finite ion temperature is negligible during
the time of the simulation.}.

The resulting time evolution of the ionic temperature for a plasma of
$100000$ ions with $\rho_{\rm i0} =2\times 10^7\:$cm$^{-3}$,
$\Gamma_{\rm{e0}}=0.08$ and $\beta=0.15\:\omega_{\rm{p,i}}(t=0)$
is shown in figure
\ref{fig3}. In the static case (dashed line in figure \ref{fig3}), the
temperature shows an initial linear rise with a slope corresponding to the
collisional heating rate $\gamma_{\rm eic}T_{\rm{e}}$ and then quickly saturates to a
value which is well described by equation (\ref{statictemp}) (marked by the
arrow in the figure). However, in the case of an expanding plasma (solid line
in figure \ref{fig3})
the temperature is drastically reduced. While the initial rise of the ion
temperature stays the same, the temperature starts to drop down already at a
relatively early time, which at this stage is mainly caused by adiabatic
cooling of the ions due to the plasma expansion. At later times, the ion
temperature is driven to its steady-state value given by the balance between
the collisional heating and the laser cooling (equation
(\ref{statictemp})). The long-time behaviour of the ionic temperature
can be estimated from the plasma dynamics discussed in \cite{PPR04a}, where it
was found 
that laser cooling strongly alters the expansion behaviour, leading to
an increase of the plasma width $\sigma$ according to $\sigma\stackrel{\beta
t\gg1}{\propto}t^{1/4}$. It was also shown there that the adiabatic law for the
selfsimilar plasma expansion, $\sigma^2 T_{\rm{e}}={\rm{const.}}$, still holds,
which gives together with equation (\ref{statictemp}) for $T_{\rm{c}} \to 0$
\begin{equation} \label{selfs}
T_{\rm{i}}\propto \frac{1}{\sigma^3 T_{\rm{e}}^{1/2}}\propto \sigma^{-2}\propto
t^{-1/2} \; ,
\end{equation}
while the ionic Coulomb coupling parameter increases according to
$\Gamma_{\rm{i}}\propto 1/(\sigma T_{\rm{e}})\propto t^{1/4}$.
As can be seen in the inset of figure \ref{fig3}, equation (\ref{selfs}) is
well reproduced by the numerical simulation. Hence, it is indeed this
adiabatic cooling of the expanding plasma together with a decreasing
collisional heating rate $\gamma_{\rm{eic}}T_{\rm{e}}$ which leads to much lower temperatures and
consequently to much larger Coulomb coupling parameters than those predicted
by a static estimate.

\begin{figure}[bt]
\centerline{\psfig{figure=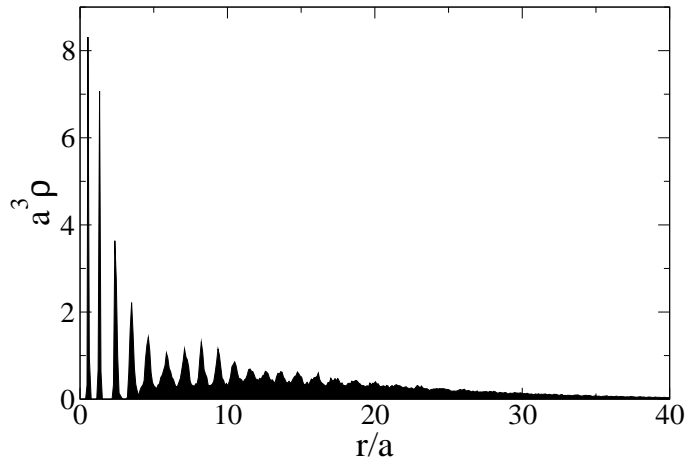,width=9cm}}
\caption{\label{fig4} Radial density of a plasma of 50000 ions with
initial-state parameters $\rho_{\rm i0}=2\times 10^7\:$cm$^{-3}$,
$\Gamma_{\rm{e0}}=0.08$, $\beta=0.17\:\omega_{\rm{p,i}}(t=0)$ and $T_{\rm{c}}=0.8\:$mK
after $t=175\:\mu$s, demonstrating the radial ordering
in the inner plasma region.}
\end{figure}
The simulations described above demonstrate that electron-ion collisions do
not prohibit achieving ionic Coulomb coupling parameters above the
crystallization limit. However,
for a given set of initial-state parameters,
$\Gamma_{\rm{i}}$ can be significantly reduced by collisional heating during
the plasma evolution, leading to an additional restriction of the initial-state parameters for which strongly coupled states can be achieved. Since it was found that the degree of spatial order in
the system sensitively depends on initial conditions such as number of ions,
electron temperature etc., one may wonder whether the long-range ordering
into shell structures described in \cite{PPR04a} is still observable in
simulations taking into account the electron-ion collisions. This is
demonstrated in figure \ref{fig4} for a plasma of $50000$ ions with a density
of $\rho_{\rm i0}=2\times 10^7\:$cm$^{-3}$ and $\Gamma_{\rm{e0}}=0.08$,
cooled with a damping rate of $0.17\:\omega_{\rm{p,i}}(t=0)$ and a limiting
temperature of $T_{\rm{c}}=0.8\:$mK.
The ordering into concentric shells is clearly visible in the radial density
after a time of $t=175\:\mu$s.

In summary, we have presented simulations for laser-cooled ultracold neutral
plasmas including elastic electron-ion collisions. It is found that
these collisions can have a significant influence on the final ionic
temperature, and hence the degree of correlation, achievable under given
experimental conditions. Yet, with a proper choice of initial conditions
rather large Coulomb coupling parameters can still be achieved. As we
have shown, this is due to the fact that the expansion of the plasma
drastically reduces the ion
temperature, due to adiabatic cooling of the ions as well as a decreasing
heating rate during the course of the plasma evolution. Therefore, the
resulting Coulomb coupling parameters are much larger than one would expect
from static estimates as in \cite{Kuz02b,Kil03}. We thus
conclude that our predictions made in \cite{PPR04a} about the possibility to
observe Coulomb crystallization in such a system remain valid. In particular,
we have shown here that the electron-ion collisions do not preclude the
build-up of long-range order and the formation of the shell structures
predicted in \cite{PPR04a}. The experimental realizability of the discussed
scheme has been demonstrated recently in \cite{Kil04}. There, the ion
temperature of an  ultracold Strontium plasma has been measured by driving the
core transition of the plasma ions during the gas expansion.
Since the same transition which allowed for Doppler imaging can
be used to cool the ions, these experiments provide the first step towards
a realization of strongly coupled ultracold plasmas in the laboratory. 

We would like to thank T.C.\ Killian and F.\ Robicheaux for valuable
discussions and for drawing our attention to the importance of
electron-ion collisions. Financial support from the DFG through grant RO1157/4
is gratefully acknowledged.


\section*{References}
\begin{thebibliography}{13}
\bibitem{Kil99} Killian TC, Kulin S, Bergeson SD, Orozco LA, Orzel C and
Rolston SL 1999 \PRL {\bf 83} 4776

\bibitem{Rob00} Robinson MP, Tolra BL, Noel MW, Gallagher TF and Pillet P
2000 \PRL {\bf 85} 4466

\bibitem{Rai01} Dutta SK, Feldbaum D, Walz-Flannigan A, Guest JR and
Raithel G 2001 \PRL {\bf 86} 3993

\bibitem{Mur01} Murillo MS 2001 \PRL {\bf 87} 115003

\bibitem{Rob02} Robicheaux F and Hanson JD 2002 \PRL {\bf 88} 055002

\bibitem{Kuz02} Kuzmin SG and O'Neil TM 2002 \PRL {\bf 88} 065003

\bibitem{Maz02} Mazevet S, Collins LA and Kress JD 2002 \PRL {\bf 88} 055001

\bibitem{Tka01} Tkachev AN and Yakovlenko SI 2001 {\it Quantum Electronics}
{\bf 31} 1084

\bibitem{PPR03} Pohl T, Pattard T and Rost JM 2003 \PR A {\bf 68} 010703(R)

\bibitem{Kil04} Simien CE, Chen YC, Gupta P, Laha S, Martinez YN,
Mickelson PG, Nagel SB and Killian TC 2004 \PRL {\bf 92} 143001

\bibitem{Ger03} Gericke DO and Murillo MS 2003 {\it Contrib.\ Plasma Phys.}
{\bf 43} 298

\bibitem{Kuz02b} Kuzmin SG and O'Neil TM 2002 {\it Phys.\ Plasmas} {\bf 9} 3743

\bibitem{Kil03} Killian TC, Ashoka VS, Gupta P, Laha S, Nagel SB, Simien CE,
Kulin S, Rolston SL and Bergeson SD 2003 \JPA {\bf 36} 6077

\bibitem{PPR04a} Pohl T, Pattard T and Rost JM 2004 \PRL {\bf 92} 155003

\bibitem{PPR04b} Pohl T, Pattard T and Rost JM 2004 \jpb {\bf 37} L183

\bibitem{PPR04c} Pohl T, Pattard T and Rost JM 2004 \PR A {\bf 70} 033416

\bibitem{Kin66} King IR 1966 {\it Astron.\ J.} {\bf 71} 64

\bibitem{Rob03} Robicheaux F and Hanson JD 2003 {\it Phys.\ Plasmas} {\bf 10}
2217

\bibitem{Met99} Metcalf HJ and van der Straten P 1999 {\it Laser Cooling and
Trapping} (Springer, New York)

\bibitem{Nan80} Nanbu K 1980 {\it J.\ Phys.\ Soc.\ Jpn.} {\bf 49} 2042

\bibitem{Fri98} Friedrich H 1998 {\it Theoretical Atomic Physics} (Springer,
Berlin) 

\bibitem{Li01} Li D 2001 {\it Nucl.\ Fusion} {\bf 41} 631

\bibitem{unpub} Pohl T 2004 {\it PhD Thesis}, TU Dresden, Germany
(in preparation)

\bibitem{Dub99} Dubin DHE and O'Neil TM 1999 \RMP {\bf 71} 87

\bibitem{Ng74} Ng KC 1974 {\it J.\ Chem.\ Phys.} {\bf 61} 2680

\bibitem{Ham96} Hamaguchi S, Farouki RT and Dubin DHE 1996 {\it J.\ Chem.\
Phys.} {\bf 105} 7641

\bibitem{Ger02} Gericke DO, Murillo MS and  Schlanges M 2002 \PR E {\bf 65}
036418

\bibitem{Lee84} Lee YT and More RM 1984 {\it Phys.\ Fluids} {\bf 27} 1273

\bibitem{Dha98} Dharma-wardana MWC and Perrot F 1998 \PR E {\bf 58} 3705;
erratum 2001 {\it ibid.} {\bf 63} 069901

\bibitem{Haz01} Hazak G, Zinamon Z, Rosenfeld Y and Dharma-wardana MWC 2001
\PR E {\bf 64} 066411
\end{thebibliography}
\end{document}